\documentstyle[12pt, epsfig,euscript,amsfonts, amsmath,amssymb,graphicx]{article}

\newcommand{\vect}[1]{{\boldsymbol #1}}

\title{Discrete Time Leads to Quantum-Like Interference of Deterministic Particles}

\author{Andrei Khrennikov\\ 
International Center for Mathematical Modeling\\
in Physics, Engineering and Cognitive science\\
MSI, V\"axj\"o University, S-35195, Sweden\\
email: Andrei.Khrennikov@msi.vxu.se\\
Yaroslav Volovich\thanks{Project of The Royal Academy of Science, Sweden,
on the collaboration with States of the former Soviet Union; and Project
"Mathematical Modelling" of V\"axj\"o University.}\\
Physics Department, Moscow State University\\
Vorobievi Gori, 119899 Moscow, Russia\\
email: yaroslav@aylabs.com}

\begin{document}

\maketitle

\begin{abstract}
In this note we demonstrate that a quantum-like interference picture could appear
as a statistical effect of interference of deterministic particles, i.e. particles
that have trajectories and obey deterministic equations, if one introduces a discrete
time. The nature of the resulting interference picture does not follow from the geometry
of force field, but is strongly attached to the time discreetness parameter.
As a demonstration of this concept we consider a scattering of charged particles on
the charged screen with a single slit. The resulting interference picture
has a nontrivial minimum-maximum distribution which vanishes as the time discreetness
parameter goes to zero that could be interpreted as an analog of quantum decoherence.
\end{abstract}

\section{Introduction}

It is well known that historically the results of 
experiments with elementary particles were interpreted
as the evidence of the impossibility to provide {\it deterministic,}
classical-like, description of motion of these objects. The main attitude
in the development of quantum theory was deeper and deeper understanding 
of the fact that {\it quantum randomness} has fundamental, irreducible,
character -- in the opposite to {\it classical randomness.} 
Randomness in classical statistical mechanics can be reduced to 
uncertainty of initial conditions. The evolution of 
probabilistic density described by Liouville's equation can be reduced 
to deterministic evolution of an ensemble of particles described by Hamiltonian
equations on the phase space. As we have just mentioned, such a picture was considered
as totally inadequate to quantum situation. The formulation by N. Bohr
the {\it complementarity principle} was the culmination of anti-deterministic development
of views to experiments with elementary particles. The collection of these views
(originated by Bohr, Heisenberg and many others, see e.g. [1], [2]) is now days
known as the {\it orthodox Copenhagen interpretation} of quantum mechanics.
By this interpretation it is (even in principle) impossible to provide deterministic
description of motion of elementary particles. Thus we could not 
reproduce statistical results of quantum experiments by using classical statistical
mechanical approach: deterministic equations for trajectories of e.g. electron
that reproduce statistical behaviour described by a wave function. This
idea appeared already in letter's exchange between Heisenberg and Bohr directly
after the publication of 
the famous Heisenberg paper [3]. By discovering dynamical equations that describe
physical observables, e.g. position and momentum observables, W. Heisenberg claimed
that deeper, ontic, description of quantum systems is even in principle impossible.
This viewpoint was strongly supported by the discovery of Heisenberg's {\it uncertainty
principle.} This principle was used by N. Bohr as the basis for starting great 
changes in philosophy of physics, resulted by the complementarity principle.

We would like to notice, see [4] for the details, that, in fact,  Heisenberg-Bohr
conclusions were not logically justified. By creating a mathematical
formalism for physical observables, Heisenberg matrice-mechanics, we do not 
prove the impossibility of {\it ontic deterministic description.} Neither by referring
to uncertainty principle. If this principle would be interpreted statistically,
see e.g. [5], [6], then this would be simply a relation for
dispersions of two random  variables. It seems very doubtful that such a relation
should imply such strong restrictions on mathematical description of 
reality as e.g. impossibility to describe trajectories of e.g. electrons in configuration
or phase space. 

Now days we can definitely say that Heisenberg-Bohr conclusions
were not totally justified, since we have e.g. {\it Bohmian formalism} [7], [8] or 
{\it stochastic electrodynamics,} see e.g. [9], [10] on results related to this paper.
For example, the  Bohmian formalism provides the deterministic description of trajectories
of e.g. electron. Statistical results given by quantum formalism can be reproduced
on the basis of this deterministic picture. However, by some reasons Bohmian 
theory is commonly considered as unacceptable. It seems that the appearance
of this ontic model for quantum mechanics does not change essentially
the orthodox Copenhagen orientation of quantum community. By ourself we do not have
definite point of view to the validity of Bohmian model. In any case it could be used
as an argument against Heisenberg-Bohr conclusions. 

We think that there might be created other, non-Bohmian, ontic models
reproducing probabilistic results given by  quantum formalism. In particular,
it might be that some of these models could be local. Of course, the reader
may argue that there are {\it Bell's arguments.} These arguments imply
that local deterministic
description is impossible, see e.g. [11], [12]. However, recent investigations, see
e.g. [13] (papers of Accardy and Regoli, Ballentine,
De Muynck, Gudder, Volovich), [14] --[16], demonstrated that Bell's conclusions were not totally justified.
It seems that experimental violations of Bell's inequality need not be interpreted
as arguments against local realism. 

{\it `Local realism'} is the standard terminology used in Bell-discussions. Of course, the 
use of such a terminology was a consequence of EPR-discussions. We would not like to
tell about realism. Of course, we do not deny the existence of independent physical reality.
But we understood well that all our models are simply approximations of this reality.
We would never create the model that would be totally adequate to physical reality.
We prefer to speak about deterministic models that provide some mathematical picture
of reality. The crucial point of this consideration is that, in principle,
a mathematical model of space-time need not be identified with the {\it `continuous'
model} given by real numbers. When we say deterministic description, we do not mean
that this is some model based on differential equations in the continuous real space.
For instance, it can be some discrete {\it deterministic model} or a $p$-adic
model. The latter models were intensively studied in the connection to string
theory, see, for instance, [17]-[21]. At the moment we do not say the $p$-adic
program of reconstruction of theoretical physics was totally successful. In any case
it is an interesting (deeply developed) alternative approach to mathematical
modelling of physical phenomena. However, it is not the subject of our present consideration.

Regarding to space-time models, we, finally, remark that locality in the most of `local
realism' discussions also is considered in the old fashioned, real continuous, form. It would be 
natural to extend this 18th-20th century approach and consider not only real locality,
but e.g. locality on a discrete space or $p$-adic locality. As having large
experience of the work with $p$-adic numbers, we can tell that there is crucial
difference between e.g. real and $p$-adic locality.

As the reader understood, we are looking for local deterministic models in above mentioned
extended meaning that could reproduce probabilistic behaviour described by 
quantum theory. 

We recall that one of the most distinguishing features of 
`quantum' probabilistic behaviour is the appearance of {\it interference} structures
created by ensembles of elementary particles. In fact, the impossibility to provide
deterministic description of such a phenomenon was the main reason to create
the orthodox Copenhagen interpretation. Thus deterministic models reproducing
interference pictures are of the large interest as at least simulating
quantum behaviour. 

We remark that according to the orthodox Copenhagen interpretation it is (even in principle)
impossible to create such models. In particular, we can not describe `self-interference'
of e.g. electron in the two slit experiment without to use wavelike arguments. In particular,
there is a rather common opinion that there is crucial difference between classical and quantum
probabilistic rules for addition of probabilities of alternatives:
\begin{equation}
\label{F1}
P=P_1+P_2
\end{equation}
\begin{equation}
\label{F2}
P=P_1+P_2+2\sqrt{P_1P_2}\cos\theta.
\end{equation}
However, recently it was demonstrated in series [22]-[24] of papers that 
the difference between `classical' and `quantum' probabilities
can be explained in classical probabilistic terms by taking into account
{\it contextual dependence of probabilities} involved into quantum interference.
Moreover, in [22]--[24] there was presented the idea that quantum formalism
was merely the pure mathematical discovery of such calculus of probabilities
depending on contexts, complexes of experimental physical conditions.
Of course, everybody knows that contextualism was originally incorporated
into quantum theory by N. Bohr. Thus, in fact, contextual probabilistic
approach is nothing else than mathematical (classical probabilistic)
formalization of {\it Bohr's contextualism.} We also remark that 
`classical probabilistic' should not identified with some concrete
probabilistic model; in particular, conventional measure-theoretical model,
Kolmogorov, 1933 [25]. For us, `classical probabilistic' is  a frequency description.
There can be various mathematical models for such a description. Of course, quantum
probabilistic calculus, Hilbert space calculus, also gives a frequency description.
However, here we start with Hilbert space that appears without direct relation
to frequencies; such a Hilbert space description (with corresponding interpretation
of superposition principle) looks merely as the mathematical model for wave 
phenomena. In [22]--[24] we start with frequencies and reproduce  complex
wave amplitudes and the Hilbert space structure as a consequence of contextual
dependence of probabilities.

Understanding of the contextual structure of quantum probabilities
implies that, on one hand, we need not follow to orthodox Copenhagen
\footnote{In particular, there can be created local deterministic models for
quantum statistics.}; on the  other hand, quantum-like  probabilistic
behaviour need not be related only to experiments with elementary 
particles. In particular, we can try to obtain interference-like effects
in {\it experiments with macroscopic systems} by taking into account context dependence.
In our paper [26] we presented numerical experiment for macroscopic charged balls
that can be considered as
the direct analogue of the two slit experiment. We found the interference effect and
introduced corresponding complex waves (of course, of probabilities).

In this paper we study the deterministic model for
scattering of charged particles on
the charged screen with a single slit. The resulting interference picture
has a nontrivial minimum-maximum distribution. This interference picture
has no relation to `self-interference' of particles, no wave-structure 
is involved into considerations. The basic source of interference is the discrete
time scale used in our mathematical model: instead of Newton's differential
equations, continuous time evolution, we consider difference equations,
discrete time evolution. Interference effect disappears as the time discreetness
parameter goes to zero that could be interpreted as an analog of quantum decoherence.

The common viewpoint might be that we study just a discrete approximation 
to continuous Newton model. It is supposed that the latter model gives the 
right picture of `classical physical reality.' However, we think
that {\it continuous Newton's model is just an approximation of physical reality.}
The right picture is given by discrete difference equations. Hence the 
contradiction between statistical description provided by quantum formalism and
Newtonian mechanics could not be considered as a contradiction between
quantum and classical (deterministic) physics. Such a contradiction, that typically
discussed in quantum literature,  should be interpreted  as simply a consequence
of the use of an approximation, namely continuous Newtonian mechanics, instead 
of the use of the adequate model, namely discrete model with some level
of discretization depending on an experimental context. 

\section{Discrete Time in Newton's Equations}

Classical particles are believed to obey the well known Newton's equation
\begin{equation}
\label{eq}
\vect{F}=m\ddot{\vect{r}}
\end{equation}
Here we modify this equation to produce an interference picture similar to quantum
interference. We introduce a parameter of time discreetness $\tau$ described below.

Let us rewrite the second order differential equation (\ref{eq}) as a system of
first order differential equations, we have
\begin{equation}
\label{e1}
\begin{split}
\vect{F}&=m\dot{v}\\
v&=\dot{r}
\end{split}
\end{equation}
In the system (\ref{e1}) the derivatives assume the continuousness of time.
Let us now introduce a discreetness parameter $\tau$.
\begin{equation}
\label{e2}
\begin{split}
\vect{F}&=m \frac{v(t+\tau)-v(t)}{\tau}\\
v(t+\tau)&=\frac{r(t+\tau)-r(t)}{\tau}
\end{split}
\end{equation}
In the limit of $\tau\to 0$ (\ref{e2}) is equivalent to (\ref{eq}) and (\ref{e1}).

In the model described below we consider particles which move obeying the system
(\ref{e2}) where the force is produced by a charged screen.

Please note that in our model the coordinate space is left continuous, although
it would be interesting to consider it on the discrete coordinate space, i.e.
on the lattice.


\section{The Model}

\begin{figure}[htbp]
  \begin{center}
    \epsfig{file=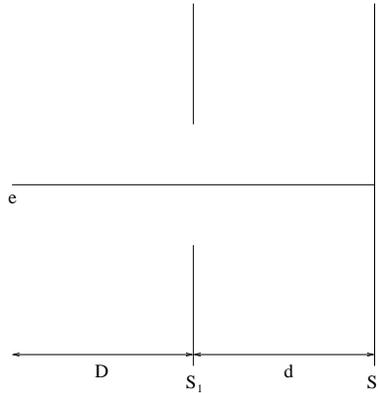,width=5cm}
\caption{Single slit experiment. Charged particles are emitted at point e pass through
a slit in the screen $S_1$ and gather on the screen $S_2$.}
\end{center}
\end{figure}


We consider a scattering on the single slit (Fig.1).
Uniformly charged round particles are emitted at point $e$ (emitter) with fixed
velocity with angles evenly distributed in the range $(-\pi/2,\pi/2)$. Each
particle interacts with the uniformly charged flat screen $S_1$. The charge
distribution on the particle and the screen stays unchanged even if the
particle comes close to the screen. Physically this is a good approximation
when the particle and the screen are both made of dielectric. There is a
rectangular slit in the screen (on the Fig.1 the slit is
perpendicular to the plane of the picture). Particles pass through the
slit in screen $S_1$ and gather on screen $S_2$. We are interested in the particle
distribution on the second screen.

Now let us write the laws of motion for the particles.
The force affecting the particle is given by the Coulomb's law
\begin{equation}
\label{ex}
\vect{F_i}=\int\limits_{\mathcal{D}_i}
           \frac{q\sigma}{|\vect{r'}|^2}\cdot\frac{\vect{r'}}{|\vect{r'}|}~
		   ds
\end{equation}
where $\vect{r'}$ is a vector from an element on the screen to the
particle, $q$ is charge of the particle, $\sigma$ is charge density on the
screen, i.e. charge of a unit square. We integrate over the surface of the
screen, the integration region $\mathcal{D}_i$ is plane of the screen
except the split.

Projecting equation (\ref{ex}) to $xy$-plane, where $x$ and $y$
denotes horizontal and vertical coordinates of the particle respectively we get
\begin{equation}
\label{em}
\begin{split}
F_x &= q\sigma \int\limits_{\Gamma} dy' \int\limits_{\mathbb{R}}
dz' \frac{x}{(x^2+(y-y')^2+{z'}^2)^{3/2}}\\
F_y &= q\sigma \int\limits_{\Gamma} dy' \int\limits_{\mathbb{R}}
dz' \frac{y-y'}{(x^2+(y-y')^2+{z'}^2)^{3/2}}
\end{split}
\end{equation}
where $2R$ is the height of the slit, $F_x$ and $F_y$ denote the projections
of the force $\vect{F}$ to $x$ and $y$ axes, and the integration region
\begin{equation}
\label{eg}
\Gamma=(-\infty,-R)\cup(R,+\infty)
\end{equation}

Integrating the rhs of (\ref{em}) we get
\begin{equation}
\label{force}
\begin{split}
F_x &= 2q\sigma\left(\pi+\arctan{\frac{y-R}{x}}-\arctan{\frac{y+R}{x}}\right)\\
F_y &= q\sigma\ln\frac{x^2+(R-y)^2}{x^2+(R+y)^2}
\end{split}
\end{equation}

We take the following initial values
\begin{equation}
\label{ic}
\begin{aligned}
x(0) &=-D\\
y(0) &= 0
\end{aligned}
\qquad
\begin{aligned}
\dot{x}(0) = v_0\cos\alpha\\
\dot{y}(0) = v_0\sin\alpha
\end{aligned}
\end{equation}
where angle $\alpha$ is a random variable uniformly distributed in
$[0,2\pi)$. The constant parameters $v_0$ and $D$ are initial velocity and
distance between emitter and the screen.

Particles are emitted at point $e$ (see Fig.1), move obeying
affected by force (\ref{force}) pass through slit in the screen $S_1$ and
gather on the screen $S_2$. Having points where particles hit the screen
$S_2$ we compute frequencies with which particles appear on screen $S_2$ as
a function of coordinates on the screen, we call this function a particle distribution.
We are interested in computing the particle distribution over a vertical line on
screen $S_2$ with $z=0$. That is why we consider a motion only in the
$xy$-plane and initial values (\ref{ic}) do not contain $z$-coordinate.

The second screen was separated with cells of equal size, the diameter of a particle.
The number of particles which hit into each cell was calculated and interpreted
as a particle distribution.
The details of numeric computations are given in the appendix.

\section{Conclusion}

In this note we have shown that a quantum-like interference picture could appear
as a statistical effect of deterministic particles, i.e. having trajectories and obeying
deterministic equations, if one introduces a discrete time. The nature of the resulting
interference picture (particle distribution, see Fig. 2-5 in appendix) does not
follow from the geometry of force field, but is strongly attached to the discreetness
parameter $\tau$.

The described behavior stays without contradiction with a contextual approach to
quantum probabilities. It would be interesting to investigate the scattering on the
two slit screen.

\section{Acknowledgments}

One of the authors, Andrei Khrennikov, would like to thank 
I would like to thank 
S. Albeverio, L. Accardi, L. Ballentine, V. Belavkin, E. Beltrametti, W. De Muynck,
R. Gill, D. Greenberger, S. Gudder, T. Hida, A. Holevo,  P. Lahti, D. Mermin,  A. Peres, J. Summhammer,  
for (sometimes extremely critical) discussions on probabilistic foundations of quantum 
mechanics.

\medskip

{\bf References}

\medskip

1.  N. Bohr, {\it Phys. Rev.,} {\bf 48}, 696-702 (1935).

2. W. Heisenberg, {\it Physical principles of quantum theory.}
Chicago Univ. Press, 1930.

3. W. Heisenberg, {\it Zeits. f\"ur Physik,} {\bf 33},  879 (1925).

4. A. Khrennikov, {\it V\"axj\"o interpretation of quantum mechanics.} Preprint 
quant-ph/0202197.

5. H. Margenau, {\it Phil. Sci.,} {\bf 25}, 23 (1958).

6. L. E. Ballentine,  The statistical interpretation of quantum mechanics,
{\it Rev. Mod. Phys.,} {\bf 42}, 358--381 (1970). 

7. D. Bohm (1951), {\it Quantum theory, Prentice-Hall.} 
Englewood Cliffs, New-Jersey.
 
8.  D. Bohm  and B. Hiley,  {\it The undivided universe:
an ontological interpretation of quantum mechanics.} Routledge and Kegan Paul, 
London, 1993.

9. G. Cavalleri, {\it Nuovo Cimento} B, {\bf 112},  1193 (1997) .

10. A. Zecca   and G. Cavalleri , {\it Nuovo Cimento} B, {\bf 112},  1, (1997);
Cavalleri G.  and Tonni E., "Discriminating between QM and SED with spin",
in C. Carola and A. Rossi,
{\it The Foundations of Quantum Mechanics (Hystorical Analysis and Open Questions)}
(World Sceintific Publ., Singapore), p.111, 2000.

11. J.S. Bell,  Rev. Mod. Phys., {\bf 38}, 447--452 (1966).
J. S. Bell, {\it Speakable and unspeakable in quantum mechanics.}
Cambridge Univ. Press, 1987.

12. J.F. Clauser , M.A. Horne, A. Shimony, R. A. Holt,
Phys. Rev. Letters, {\bf 49}, 1804-1806 (1969);
J.F. Clauser ,  A. Shimony,  Rep. Progr.Phys.,
{\bf 41} 1881-1901 (1978).
 A. Aspect,  J. Dalibard,  G. Roger, 
Phys. Rev. Lett., {\bf 49}, 1804-1807 (1982);
 D. Home,  F. Selleri, Nuovo Cim. Rivista, {\bf 14},
2--176 (1991). H. P. Stapp, Phys. Rev., D, {\bf 3}, 1303-1320 (1971);
P.H. Eberhard, Il Nuovo Cimento, B, {\bf 38}, N.1, 75-80(1977); Phys. Rev. Letters,
{\bf 49}, 1474-1477 (1982);
A. Peres,  Am. J. of Physics, {\bf 46}, 745-750 (1978).
P. H. Eberhard,  Il Nuovo Cimento, B,
{\bf 46}, N.2, 392-419 (1978); J. Jarrett, Noûs, {\bf 18},
569 (1984).

13. {\it Proc. Conf. "Foundations of Probability and Physics",}
V\"axj\"o-2000, Sweden; editor A. Khrennikov, WSP, Singapore, 2001.

14. I. Pitowsky,  Phys. Rev. Lett, {\bf 48}, N.10, 1299-1302 (1982);
Phys. Rev. D, {\bf 27}, N.10, 2316-2326 (1983);
S.P. Gudder,  J. Math Phys., {\bf 25}, 2397- 2401 (1984);
A. Fine,  Phys. Rev. Letters, {\bf 48}, 291--295 (1982);
P. Rastal, Found. Phys., {\bf 13}, 555 (1983).
W. Muckenheim,  Phys. Reports, {\bf 133}, 338--401 (1986);
W. De Baere,  Lett. Nuovo Cimento, {\bf 39}, 234-238 (1984);
{\bf 25}, 2397- 2401 (1984). 

15. A. Yu. Khrennikov, {\it Interpretations of Probability.}
VSP Int. Sc. Publishers, Utrecht, 1999.

16.  A. Yu. Khrennikov, A perturbation of CHSH inequality induced by fluctuations 
of ensemble distributions. {\it J. of Math. Physics}, {\bf 41},  5934-5944(2000).

17. V. S. Vladimirov,  I.  V. Volovich, and  E. I. Zelenov, 
{\it $p$-adic Analysis and  Mathematical Physics}, WSP, Singapore, 1994.

18. P. G. O. Freund  and E. Witten, Adelic string amplitudes.
                    {\it Phys. Lett. B}, {\bf 199}, 191-195 (1987).
                    
19. P. H. Frampton  and  Y. Okada, $p$-adic string
$N$-point function. {\it Phys. Rev. Lett. B }, {\bf 60}, 484--486 (1988).

20. A. Yu. Khrennikov, {\it $p$-adic valued distributions in 
mathematical physics.} Kluwer Acad. Publ., Dordrecht, 1994.

21. A. Yu. Khrennikov, {\it Non-Archimedean analysis: quantum
paradoxes, dynamical systems and biological models.}
Kluwer Acad. Publ., Dordreht,  1997.

22. A. Yu. Khrennikov, Linear representations of probabilistic transformations 
induced by context transitions. {\it J. Phys.A: Math. Gen.,} {\bf 34}, 9965-9981 (2001).

23. A. Yu. Khrennikov, Origin of quantum probabilities. {\it Proc. Conf.
"Foundations of Probability and Physics",} V\"axj\"o-2000, Sweden; editor A. Khrennikov,  
p. 180-200, WSP, Singapore (2001).

24. A. Yu. Khrennikov, {\it Ensemble fluctuations and the origin of
quantum probabilistic rule.} Rep. MSI, V\"axj\"o Univ., {\bf 90}, October (2000).

25. A. N. Kolmogoroff, {\it Grundbegriffe der Wahrscheinlichkeitsrechnung.}
Springer Verlag, Berlin (1933); reprinted:
{\it Foundations of the Probability Theory}. 
Chelsea Publ. Comp., New York (1956).

26. A. Yu. Khrennikov,  Ya. I. Volovich, {\it Numerical experiment on interference
for macroscopic particles.} Preprint quant-ph/0111159, 2000.

\newpage

\section*{Appendix}

Below we present a sample trajectories plots for two different discretness parameters
$\tau$ (Fig.4,5). And the corresponding particle distributions (Fig.2,3).

Parameters of the model where the following $D=5, d=25, R=5, v_0=12$ the particle
and the screen $S_1$ (Fig.1) had the opposite charges, i.e. $q\sigma=-1$ and the radius
of a particle $r=0.2$.

To produce a particle distribution pictures with a trustable precision about
$10^7$-$10^8$ trajectories where computed. To produce such a large amout of computations
even for modern stations we used a parallel Sun-UltraSPARC 4-processor station located
at V\"axj\"o University.
The program was implemented using GNU C++ (g++).

Since the computation time is proportional to $1/\tau$ and the computations where performed
simultaneously the total number of computed trajectories for $\tau=0.05$ (Fig.4) and
$\tau=0.01$ (Fig.3) differs approximately five times.

\begin{figure}[htbp]
  \begin{center}
    \epsfig{file=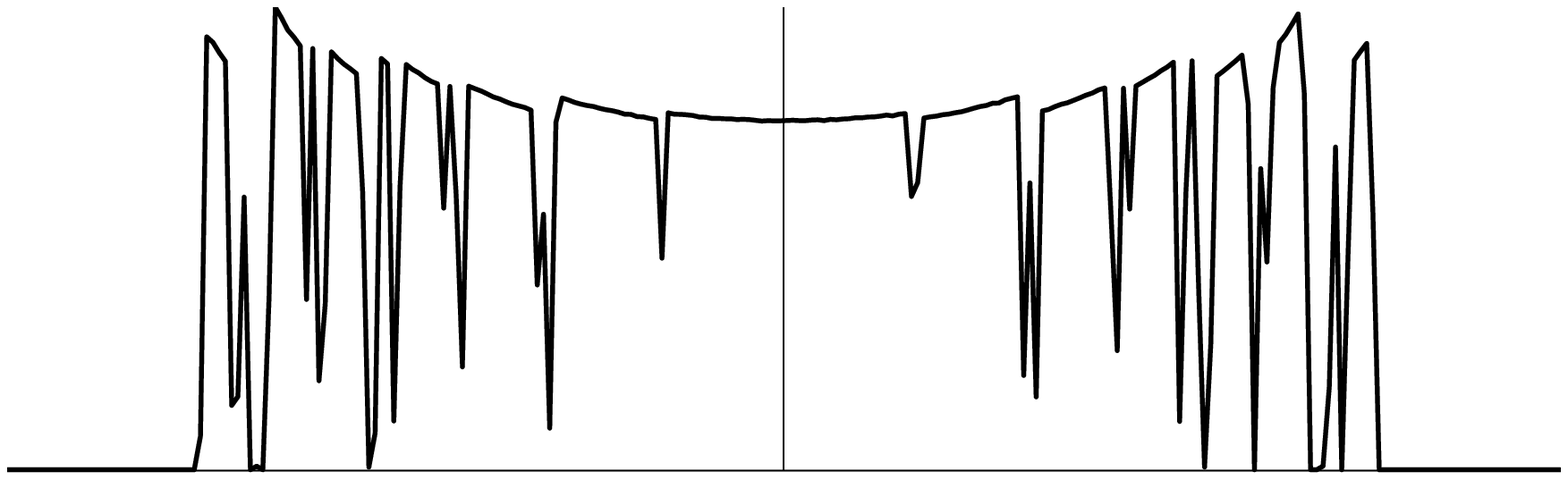,width=10cm}
\caption{Particle distribution on the second screen. See also the corresponding (Fig.4).
Parameters: $\tau=0.05$, $total=138 582 362$ particle trajectories where computed.}
\end{center}
\end{figure}


\begin{figure}[htbp]
\begin{center}
\epsfig{file=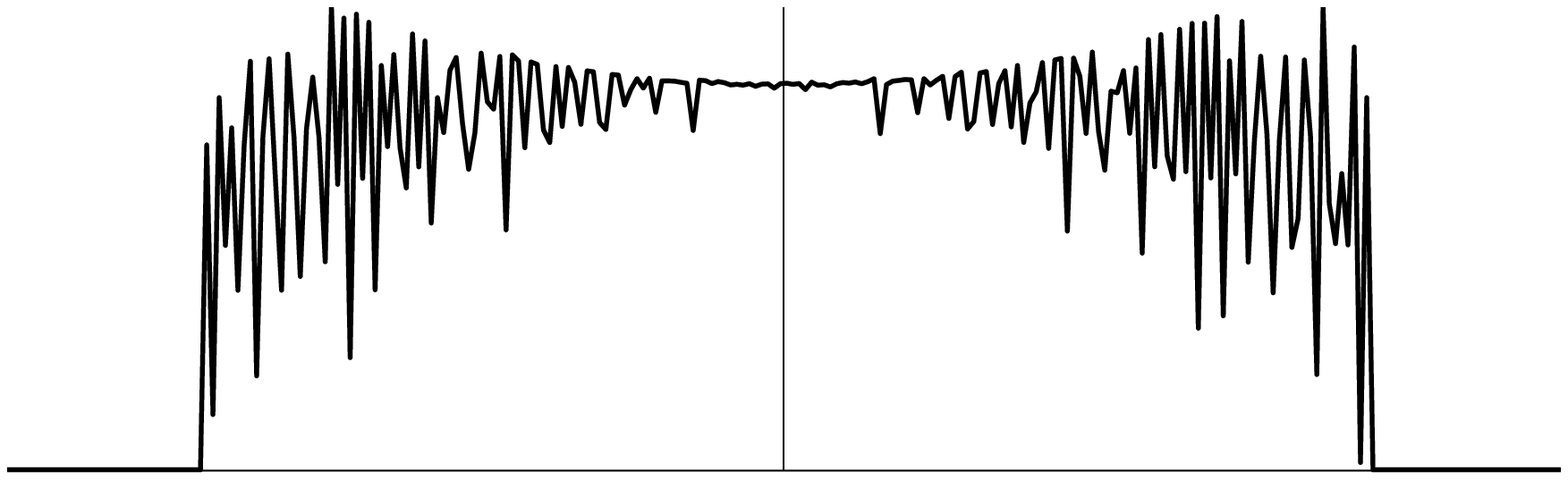,width=10cm}
\end{center}
\caption{Particle distribution on the second screen. See also the corresponding (Fig.5).
Parameters: $\tau=0.01$, $total=28 200 885$ particle trajectories where computed.}
\end{figure}

\begin{figure}[htbp]
\begin{center}
\epsfig{file=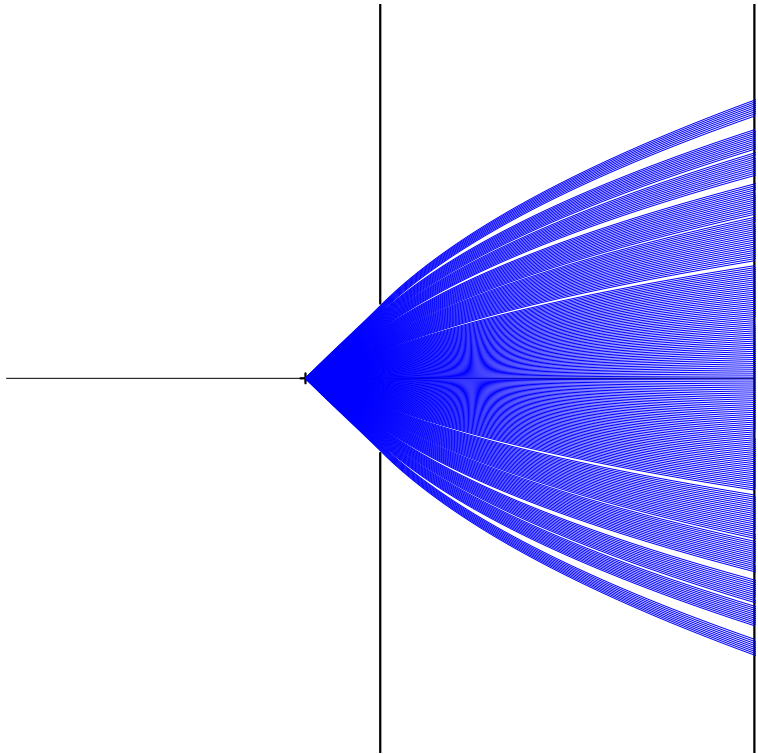,width=10cm}
\end{center}
\caption{Single slit experiment. Particles are emitted with even distribution, although
it is seen that the distribution on the second screen is nontrivial.
See the corresponding (Fig.2) where several millions trajectories were computed to
plot a distribution on the second screen. Paramteters: discretness: $\tau=0.05$,
trajectories: $250$,
angles: $-45.5 \leq\alpha\leq 45.5$}
\end{figure}

\begin{figure}[htbp]
\begin{center}
\epsfig{file=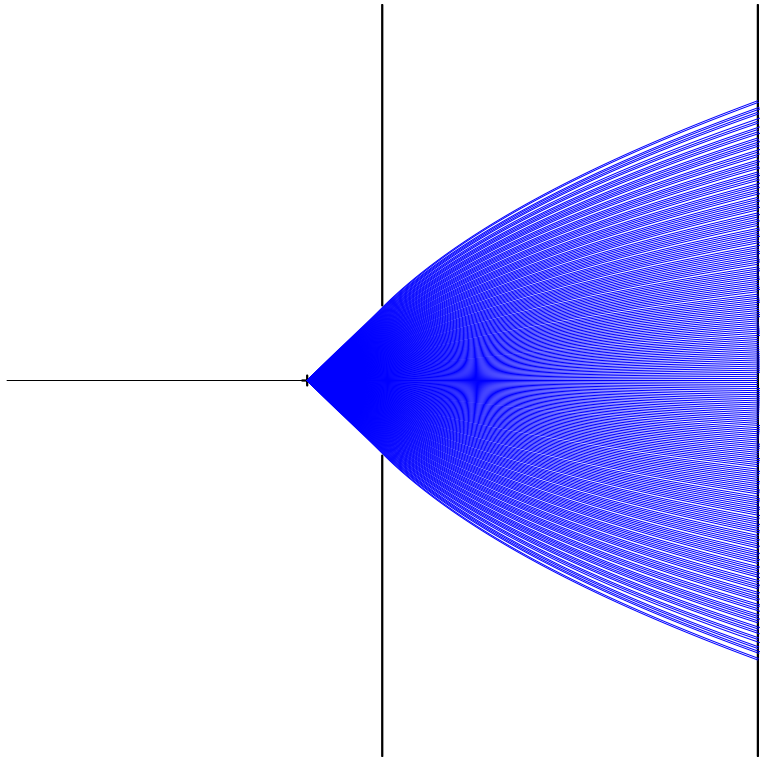,width=10cm}
\end{center}
\caption{Single slit experiment. The corresponding distribution on the second screen is
given on (Fig.3). Paramteters: discretness: $\tau=0.01$,
trajectories: $250$, angles $-45.5 \leq\alpha\leq 45.5$}
\end{figure}

\end{document}